\documentclass{sig-alternate}
\usepackage{url}
\usepackage{graphicx}
\usepackage{color,soul}
\usepackage{pgfplots}
\usepgfplotslibrary{dateplot}

\begin{document}

\title{A Study on Placement of Social Buttons in Web Pages}

\numberofauthors{2}
\author{
\alignauthor Omar Alonso \\
\affaddr{Microsoft Corp.} \\
\affaddr{Mountain View, CA} \\
\email{omalonso@microsoft.com}
\alignauthor Vasilis Kandylas \\
\affaddr{Microsoft Corp.} \\
\affaddr{Mountain View, CA} \\
\email{vkandyl@microsoft.com}
}
\date{}

\maketitle
\begin{abstract}
With the explosion of social media in the last few years, web pages nowadays include different social network buttons where users can express if they support or recommend content. Those social buttons are very visual and their presentations, along with the counters, mark the importance of the social network and the interest on the content. In this paper, we analyze the presence of four types of social buttons (Facebook, Twitter, Google+1, and LinkedIn) in a large collection of web pages that we tracked over a period of time. We report on the distribution and counts along with some characteristics per domain. Finally, we outline some research directions. \end{abstract}



\section{Introduction}

If we take a look at a random set of web pages, we can notice that many of them include sharing buttons from one or more social networking sites. These social buttons are included in a web page by content owners as a mechanism to help users share content and recommend the page. A social button consists of a visual representation that resembles an icon along with an optional counter that represents how many times a particular web page, video, article, or piece of content has been shared in a specific social networking site.  
For illustration purposes, we present a couple of examples (Figures 1 and 2):  a renowned newspaper\footnote{http://www.guardian.co.uk/football/2013/may/30/real-chelsea-juventus-international-champions-cup}, and 
an entertainment article\footnote{http://music.yahoo.com/blogs/reality-rocks/mariah-carey-not-returning-american-idol-223954657.html}. The two web pages contain the same number of buttons but the counters, presentation, and order are different. 

%\begin{figure}[h]
%\centering
%\includegraphics[width=80mm]{techcrunch.png}
%\caption{A technology article. All button counters are greater than zero.}
%\label{fig:techcrunch}
%\end{figure}

\begin{figure}[h]
\centering
\includegraphics[width=80mm]{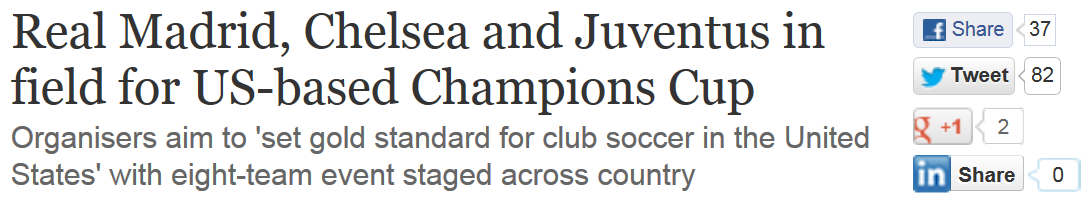}
\caption{News about a sport event. The LinkedIn button shows no activity which is somewhat expected as sports tend to be more popular in other
social networks.}
\label{fig:techcrunch}
\end{figure}

\begin{figure}[h]
\centering
\includegraphics[width=80mm]{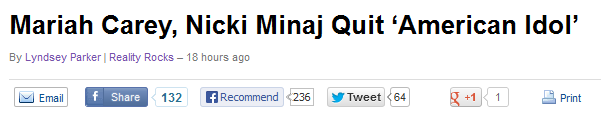}
\caption{An entertainment news article. Facebook dominates the activity with respect to counters. Also, there are two flavors of Facebook buttons
(Share and Recommend).  }
\label{fig:american-idol}
\end{figure}

Buttons are usually displayed very close to the title of the page and there is a nice layout separation to make sure the user can identify the specific 
button and read the counter correctly. Usually, buttons come in many different variations, small or large, in horizontal or vertical orientation, with or without click counters, but they all perform the same functions, to let the users show they found the web page engaging and to share it with others. The button counters, when they are shown, are important features that highlight the significance and relevance of the page. 

Studying these buttons can tell us how frequently they are used, both by web masters and by web page visitors. 
Social buttons have been around for some time in many shapes and forms and only a handful are chosen to be included in a web page. As different
social networks evolve, content owners may decide to include new or replace old ones. 

As a motivating example, say that a new social button emerges and we are interested in understanding this new phenomenon. How fast has it been adopted on the Web and how does it compare to other popular buttons?

The research questions that we pose are:

\begin{itemize}
\item	What is the distribution of social buttons in a sample of web pages?
\item	What are the most popular buttons and how do those button counters look like? 
\item	What is the social activity per domain?
\end{itemize}

To this end, we tracked a large number of web pages over several months and looked at the presence of buttons and the number of clicks that each button received for the four most common social buttons at the time: Facebook's Like, Google's +1, Twitter's Tweet and LinkedIn's Share button.
This paper is organized as follows. We first discussed related work. Second, we present our research methodology followed by a data analysis section. Finally, we present conclusions and outline some research directions.

\section{Related Work}

Counting and tracking the number of visitors to a web page has been an important metric since the early days of the Web. Showing a web counter at the bottom of the page is still a common practice in a large number of websites. With the massive adoption of social networking sites, there has been a rise on social buttons and their counters as engagement and popularity metrics. 

Very little work has been done in studying and analyzing social buttons.
Gerlitz and Helmond conducted one of the first studies on social buttons~\cite{Gerlitz11}. Their work was based on analyzing the distribution of social buttons and engagement in relation to a number of pre-defined social issues (e.g. BP Oil Spill, Tea Party, etc.). On a follow-up work, Gerlitz and Helmond emphasized the notion of the Like economy, as a framework to better understand social buttons~\cite{Gerlitz13}.  Using a data mining approach, Jin \emph{et al.}~\cite{Jin11} developed a
prototype called LikeMiner  that mines Facebook Likes and shows the influence of the social network.

In our research, we concentrate on a more large scale study that focuses on the distribution of buttons and counters over time for a set of web pages. Another contribution of this work is a framework for implementing similar studies that require a mix of machines and humans.

\section{Methodology}

We started by taking a frequency-adjusted random sample of queries from a commercial search engine log. For each query, we extracted the top 10 results from the SERP (Search Engine Result Page). This produced a set of 51,000 unique URLs, which we designate as set \emph{D1}. We also took a random sample of 10,000 URLs from \emph{D1}, which we designate as \emph{D2}.

The social networking sites provide instructions for web masters on how to include their buttons in web pages. This usually involves adding a piece of HTML or JavaScript code with some parameters that control the appearance of the button. We used the set \emph{D1} for crawling the web pages and parsed the HTML source to identify the presence of code for any of the tracked social buttons. We used a set of regular expressions to look for all possible variations that the code of the tracked can take. This process of crawling \emph{D1} and automatically looking for buttons on the web pages was repeated weekly. 

The set \emph{D2} was used to identify the number of button clicks for those buttons that had counters.This process was repeated every four weeks. The reason we used the smaller set \emph{D2} and a lower measurement frequency was because we used human computation to extract the click numbers. 
We faced several difficulties when extracting the button click counts. To get these values, a simple crawl is not enough. The web page needs to be rendered and the value correctly extracted. However, social buttons can come in many variations, with or without click counters, single- or multi-line height, etc. Web masters can customize the look and feel of buttons and, at the same time, a button can change several icons over time. Additionally, many pages show multiple buttons. For example, some blogs show one Facebook Like button for the post and a second Like button for the whole blog. The button for the whole blog has a much higher click count, but it does not reflect how many readers liked the blog post. What we want to measure is the Like count for the post. We therefore chose to use human computation to extract the click counts from the appropriate button. If a web page had more than one social button, we asked workers to report the number of clicks on each button on the page. 

Even for humans, this task is not as straightforward as one may think. The data in the counters changes from page to page. We encountered many examples where the counters had an abbreviated form, for example, ``18.9K" or were empty. Normalizing such data (K to 1,000; empty to 0) was also part of the data post-processing step. We iterated many times on the instructions for detecting official buttons and the correct counters. 
A further source of difficulties was the delay for rendering the buttons and their counters. We noticed that for many pages with a lot of buttons there is a significant delay from the moment the page loads in the browser and until the social buttons start appearing. Sometimes the buttons will appear first and then the counter would get updated. These delays meant that sometimes the human workers would report no button or a button without a counter when one would have appeared if they had waited longer.  We used an internal tool for crowdsourcing the task.

While it may be possible to reach a fully automated solution, we decided to include humans in the loop. Social buttons exist so they can generate more engagement and we believe that the ability to identify them correctly by a user is very important.
We used crowdsourcing quality control techniques (provided detailed guidelines and examples and used 3 judgments per web page) to reduce the number of disagreements in the human judgments to less than 10\%.

The study run from July 2011 till May 2012. A change in how we identified buttons automatically in the set \emph{D1} made the early data not comparable, so we omit them from the results in the next section. We also conducted calibration tests on the performance of the human workers for the first few weeks, and these are also omitted from the results for the set \emph{D2}.

\section{Data Analysis and Findings}

Each week, we monitored the presence of social buttons on the pages of the URL set \emph{D1}. Every 4 weeks we also recorded the button click counts on the pages of the URL set \emph{D2}, as was explained previously.

The study took place between July 2011 and May 2012.  As the Google +1 button was released in June 2011, the study shows how fast the new social button use was spreading across the Web and how fast it was accumulating user clicks compared to the other buttons.

\begin{figure}[h]
\centering
%\tikzstyle{every pin}=[fill=white, draw=black, font=\footnotesize]
\begin{tikzpicture}
\begin{axis}[
	xtick={2011-10-04,2011-12-22,2012-03-08,2012-05-31},
	date coordinates in=x,
	xticklabel={\year-\month-\day},
   	width=9cm,
        	%title={Button placements},
        	%xlabel=Date,
        	%ylabel={Web page count},
        	legend style={at={(0.5,1.18)}, anchor=north, legend columns=-1},
]
\addplot [color=red, mark=.] table [x=Date, y=Facebook] {button_presence.data};
\addplot [color=blue, mark=.] table [x=Date, y=Google] {button_presence.data};
\addplot [color=green, mark=.] table [x=Date, y=Twitter] {button_presence.data};
\addplot [color=purple, mark=.] table [x=Date, y=LinkedIn] {button_presence.data};
\legend{Facebook, Google, Twitter, LinkedIn}
\end{axis}
\end{tikzpicture}
\caption{Number of pages containing social buttons.}
\label{fig:button_presence}
\end{figure}
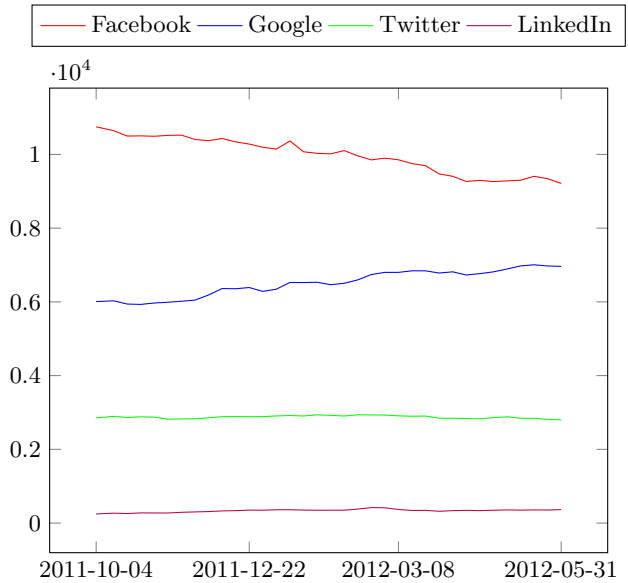

Figure 4 shows the number of web pages in the scraped search results for the 51,000 URLs in the set \emph{D1} that contain at least one of the social buttons.

As expected, the pages with Google buttons are increasing, consistent with more and more web sites starting to use the new button. Surprisingly, very soon after its introduction, the Google button had surpassed Twitter's in terms of web presence. Towards the end of the study the number of pages with Google buttons was approaching those with Facebook buttons. 

What is more surprising is that the number of pages that contained Facebook buttons showed a decrease during the time of the study. This means that some of the web pages used to include a Facebook button, but decided to remove it at a later time, but we can only speculate about the reason.
LinkedIn buttons are the least frequently occuring of the four. 

\begin{figure}[h]
\centering
%\tikzstyle{every pin}=[fill=white, draw=black, font=\footnotesize]
\begin{tikzpicture}
\begin{semilogyaxis}[
	xtick={2011-07-19,2011-10-20,2012-02-09,2012-05-03},
	date coordinates in=x,
	xticklabel={\year-\month-\day},
	width=9cm,
        	%title={Gap distribution},
        	%xlabel=Task,
        	%ylabel=Gap,
        	legend style={at={(0.5,1.11)}, anchor=north, legend columns=-1},
]
\addplot [color=red, mark=.] table [x=Date, y=Facebook] {average_clicks.data};
\addplot [color=blue, mark=.] table [x=Date, y=Google] {average_clicks.data};
\addplot [color=green, mark=.] table [x=Date, y=Twitter] {average_clicks.data};
\addplot [color=purple, mark=.] table [x=Date, y=LinkedIn] {average_clicks.data};
\legend{Facebook, Google, Twitter, LinkedIn}
\end{semilogyaxis}
\end{tikzpicture}
\caption{Average number of button clicks per page, for pages with buttons. Note that the y-axis uses a logarithmic scale.}
\label{fig:average_clicks}
\end{figure}
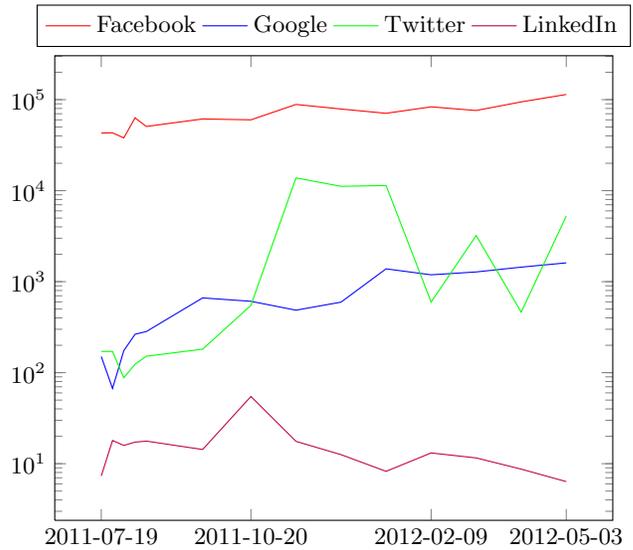

\begin{figure}[h]
\centering
%\tikzstyle{every pin}=[fill=white, draw=black, font=\footnotesize]
\begin{tikzpicture}
\begin{semilogyaxis}[
	xtick={2011-07-19,2011-10-20,2012-02-09,2012-05-03},
    	date coordinates in=x,
	xticklabel={\year-\month-\day},
    	width=9cm,
        	%title={Gap distribution},
        	%xlabel=Task,
	%ylabel=Gap,
        	legend style={at={(0.5,1.11)}, anchor=north, legend columns=-1},
]
\addplot [color=red, mark=.] table [x=Date, y=Facebook] {median_clicks.data};
\addplot [color=blue, mark=.] table [x=Date, y=Google] {median_clicks.data};
\addplot [color=green, mark=.] table [x=Date, y=Twitter] {median_clicks.data};
\addplot [color=purple, mark=.] table [x=Date, y=LinkedIn] {median_clicks.data};
\legend{Facebook, Google, Twitter, LinkedIn}
\end{semilogyaxis}
\end{tikzpicture}
\caption{Median number of button clicks per page, for pages with buttons. Note that the y-axis uses a logarithmic scale.}
\label{fig:median_clicks}
\end{figure}
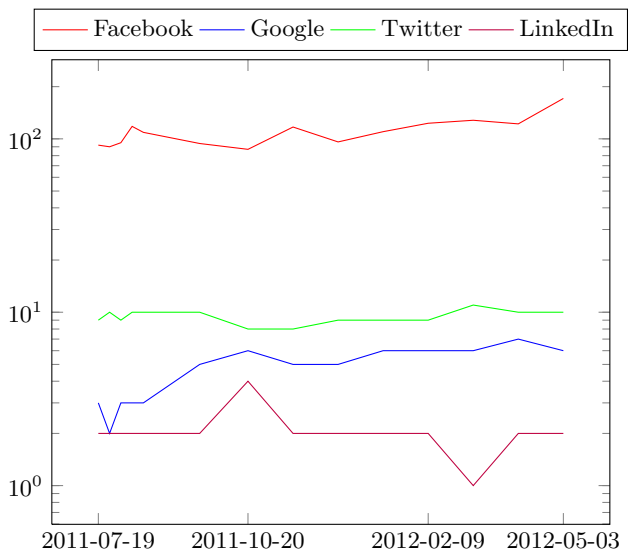

Figures 5 and 6 show the average and median number of social button clicks per page for the pages that contained buttons with click counters. These graphs are based on the subset of URLs \emph{D2} and the click values were extracted by human workers. The first five measurement points were weekly, but subsequent points were measured every four weeks. Because we use human computation, we monitored worker quality early on to make sure the data produced was accurate. After that, we decided to change the frequency of this step to every four weeks.

One would expect that the number of clicks should constantly increasing with time, but this is not the case for the charts in figures 5 and 6. There are two reasons that the average and median lines fluctuate. The first is because of errors made by the human workers. Even though we went through several iterations of revising the task and training the workers, it is impossible to eliminate all errors when dealing with humans. The second reason is that the charts show the average and median number of clicks for the web pages in set \emph{D2} that had buttons with counters. As time passed, pages that did not have a button before would add one and so they would start to be included in the computation. The newly appearing buttons though, just because they were new, would have a small number of clicks which would pull the average and median down. 

The click charts paint a different picture than the presence chart. While, as we saw in figure 4, there are more pages with Google buttons on them than pages with Twitter buttons, the Twitter buttons receive the same average but a higher median number of clicks than the Google buttons, which shows that the page visitors are more engaged with Twitter's than Google's buttons. The difference between the average and median values for the Twitter and Google buttons reveals that the distribution of clicks for the Google buttons is more uneven. A small number of buttons receive a lot of clicks, hence the average is high (about as high as Twitter's), but the median is smaller, because of the large number of buttons with few clicks.

Facebook buttons on the other hand have orders of magnitude more clicks than all the other kinds of buttons and have a clear increasing tendency in both the average and the median values. They are by far the most engaging social button for the web users.
Similar to the button presence, the clicks of the LinkedIn buttons are much lower than the other three.

As the final stage of our study, we  automatically classified the web pages containing buttons into categories  using a similar approach proposed by \cite{Bennett09}.
Table~\ref{tbl:odp_dist} shows the distribution of the URLs according to the 219 ODP (Open Dictionary Project) categories selected.
Table~\ref{tbl:odp_source} shows the top-10 domain categories for all web pages ranked by the source using Facebook prominent activity as the
ranking selection. We can observe  the commonalities between Facebook and Google+ as they share most of the categories. Twitter shares similar 
ranking at the top categories and then the categories start to change.  With LinkedIn, the rankings and categories are different as this is more of a
specific social network. LinkedIn buttons appear frequently on Name and Name Non Celebrity pages. An interesting observation is that for the category 
Science/Technology, all sources but Facebook have it on the top-10. A category like Business/Management is missing in all sources except~LinkedIn.

\begin{table}
\centering
\begin{tabular}{|l|r|} \hline
Category & Frequency \\ \hline
Arts/Writers Resources	& 13.94\% \\ \hline
Computers/Systems&	9.62\% \\ \hline
Society	&9.34\% \\ \hline
Business/Transportation and Logistics&	8.01\% \\ \hline
Uncategorized	&6.62\% \\ \hline
Health/Reproductive Health&	5.92\% \\ \hline
Science/Technology	&5.37\% \\ \hline
Shopping/Visual Arts&	4.74\% \\ \hline
Recreation/Travel	&4.39\% \\ \hline
Games/Video Games&	3.14\% \\ \hline

\end{tabular}
\caption{Distribution for the top-10 categories sorted by frequency.}
\label{tbl:odp_dist}

\end{table}

Interesting differences are the Travel and Movie-related categories that are commonly associated with Google and Facebook buttons and to a lesser degree with Twitter. The Twitter buttons also appear on Consumer Electronics and Adult pages.  Facebook buttons also appear on such pages but less frequently, whereas Google buttons are not common on name-related web pages.

\begin{table}
\centering
\begin{tabular}{|p{5cm}|r|r|r|r|} \hline
Category & FB & TW & G+ & L \\ \hline

Arts/Writers Resources & 1 & 1 & 1 & 8 \\ \hline
Computers/Systems & 2 & 2 & 2 & - \\ \hline
Society & 3 & 5 & 3 & 1 \\ \hline
Games/Video Games  & 4 & 3 & 5 & - \\ \hline
Health/Reproductive Health & 5 & 4 & 9  & 2\\ \hline
Shopping/Visual Arts & 6 & - & 5 & 5\\ \hline
Home/Gardening & 7 & 7 & 10 & -  \\ \hline
Recreation/Travel & 8 & - & - & -\\ \hline
Reference/Museums & 9 & -  & - & -\\ \hline
Business/Transportation and Logistics & 10 & 9 &4 & 3 \\ \hline
\end{tabular}
\caption{Top-10 categories ranked by source (FB~=~Facebook, TW = Twitter, G+ = Google, L = LinkedIn).}
\label{tbl:odp_source}

\end{table}

On a final note, our workers detected that a new social button (Pinterest) was gaining prominence among the set of pages. Capturing Pinterest was out of scope for our project but as anecdotal evidence, it shows that this new sharing mechanism plays an important role on how users perceive the Web nowadays. 
Social buttons are now part of the real estate of a web page. They are strategically located in the page layout with the goal of making easier for the user to recommend and share content. 

\section{Conclusions and Future Work}
In this paper, we studied and analyzed social button presence in a large data set over time and found that Facebook dominates placements and counters across the board. This is expected as Facebook is the predominant social networking site. We also noticed that the presence and counters of social buttons can differ in web domains. Part of our work included using human computation to gather better data on counters and to see if users can locate them properly. 
To the best of our knowledge, there is little research on social button placements and how the evolve over time on a corpus of web data. We shed some light into this problem and there are many more directions that we plan to continue working on.

In terms of research avenues, understanding popularity and trending content by social button activity seems a prominent area to explore more this new sharing mechanism. Analyzing placement order of buttons and replacements should be of interest as well.

\bibliographystyle{abbrv}

\begin{thebibliography}{99}

\bibitem{Bennett09}
\newblock P.~Bennett and N.~Nguyen. 
\newblock Refined Experts: Improving Classification in Large Taxonomies. In \emph{Proc. of SIGIR}. 2009. 

\bibitem{Gerlitz11}
\newblock C.~Gerlitz and A.~Helmond. 
\newblock Hit, Link, Like and Share. Organizing the social and the fabric of the web in a Like economy. \emph{DMI mini-conference} University of Amsterdam, 2011.

\bibitem{Gerlitz13}
\newblock C.~Gerlitz and A.~Helmond. 
\newblock The Like economy: Social buttons and the data-intensive web. \emph{New Media Society}. 2013.

\bibitem{Jin11}
\newblock X.~Jin and C.~Wang and J.~Luo and X.~Yu and J.~Han.
\newblock LikeMiner: a system for mining the power of 'like' in social media networks. In \emph{Proc. of KDD}. 2011.

\end{thebibliography}

\end{document}